\documentclass[%
 reprint,
 superscriptaddress,
nofootinbib,
 amsmath,amssymb,
 aps,
 pra,
]{revtex4-1}

\usepackage{graphicx}
\usepackage{dcolumn}
\usepackage{subfigure}
\usepackage{bm}
\usepackage[colorlinks=true,breaklinks=true,allcolors=blue]{hyperref}
\usepackage{url}
\usepackage{xcolor}
\usepackage{enumitem}

\makeatletter
\let\cat@comma@active\@empty
\makeatother

\begin{document}

\title{Dynamical critical scaling of long-range interacting quantum magnets}

\author{Nicol\`o Defenu}
\affiliation{Institut f\"ur Theoretische Physik, Universit\"at 
Heidelberg, D-69120 Heidelberg, Germany}
\author{Tilman Enss}
\affiliation{Institut f\"ur Theoretische Physik, Universit\"at 
Heidelberg, D-69120 Heidelberg, Germany}
\author{Michael Kastner}
\affiliation{Institute of Theoretical Physics, Department of Physics, University of Stellenbosch, Stellenbosch 7600, South Africa}
\affiliation{National Institute for Theoretical Physics (NITheP), Stellenbosch 7600, South Africa}
\author{Giovanna Morigi}
\affiliation{Theoretische Physik, Universit\"at des Saarlandes, D-66123 Saarbr\"ucken, Germany}

\date{\today}
\begin{abstract}
Slow variations (quenches) of the magnetic field across the paramagnetic-ferromagnetic phase transition of spin systems produce heat. In systems with short-range interactions the heat exhibits universal power-law scaling as a function of the quench rate, known as Kibble-Zurek scaling. In this work we analyze slow quenches of the magnetic field in the Lipkin-Meshkov-Glick (LMG) model, which describes fully connected quantum spins. We analytically determine the quantum contribution to the residual heat as a function of the quench rate $\delta$ by means of a Holstein-Primakoff expansion about the mean-field value. Unlike in the case of short-range interactions, scaling laws in the LMG model are only found for a ramp ending at the critical point. If instead the ramp is symmetric, as in the typical Kibble-Zurek scenario, after crossing the critical point the system tends to reabsorb the defects formed during the first part of the ramp: the number of excitations exhibits a crossover behavior as a function of $\delta$ and tends to a constant in the thermodynamic limit. Previous, and seemingly contradictory, theoretical studies are identified as specific limits of this dynamics.  Our results can be tested on several experimental platforms, including quantum gases and trapped ions.

\end{abstract}

\maketitle

The development of a comprehensive statistical mechanics description of out-of-equilibrium systems is a quest of relevance across disciplines, including biology, physics, computer science and financial markets \cite{chou2011}.  For long-range interactions, in particular, insights so far are mostly based on numerical simulations \cite{Campa2014}, which become increasingly involved when the dynamics is deep in the quantum regime \cite{Halimeh2017, zunkovic2018}. An open question concerns understanding the subtle interplay between time evolution, interactions, quantum, and thermal fluctuations, and specifically the connection between dynamical and equilibrium properties of quantum critical systems \cite{polkovnikov2011}. Theoretical and experimental studies of many-body critical dynamics after sudden variations of control fields have identified features which are reminiscent of the behavior of thermodynamic functions at transition points \cite{heyl2017, zhang2017}. Yet, the relation between dynamical scaling and equilibrium critical phenomena is elusive and often only conjectured. 

\begin{figure}[ht]
\subfigure[]{\includegraphics[scale=.05]{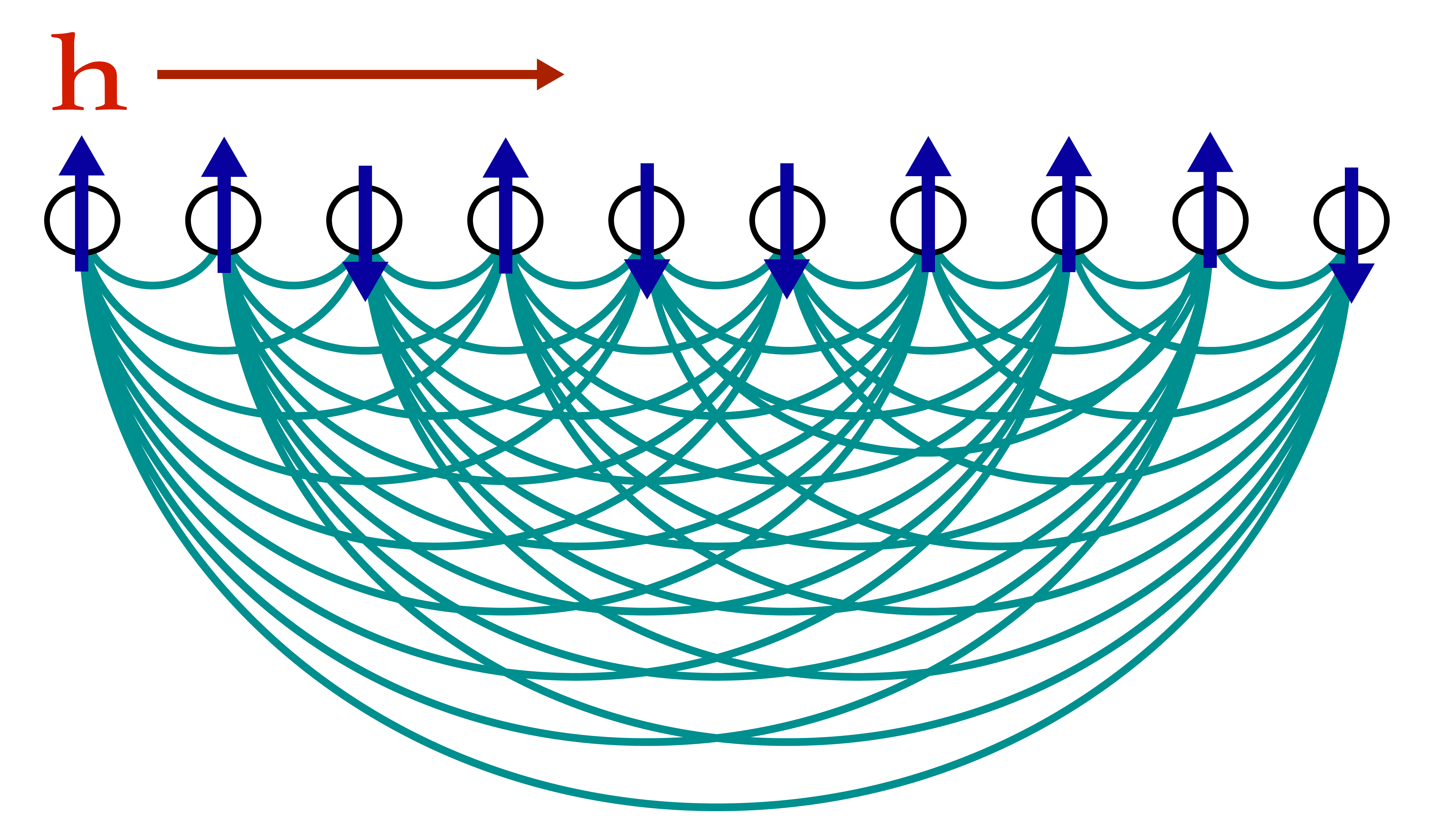}}
\subfigure[]{\includegraphics[scale=.2]{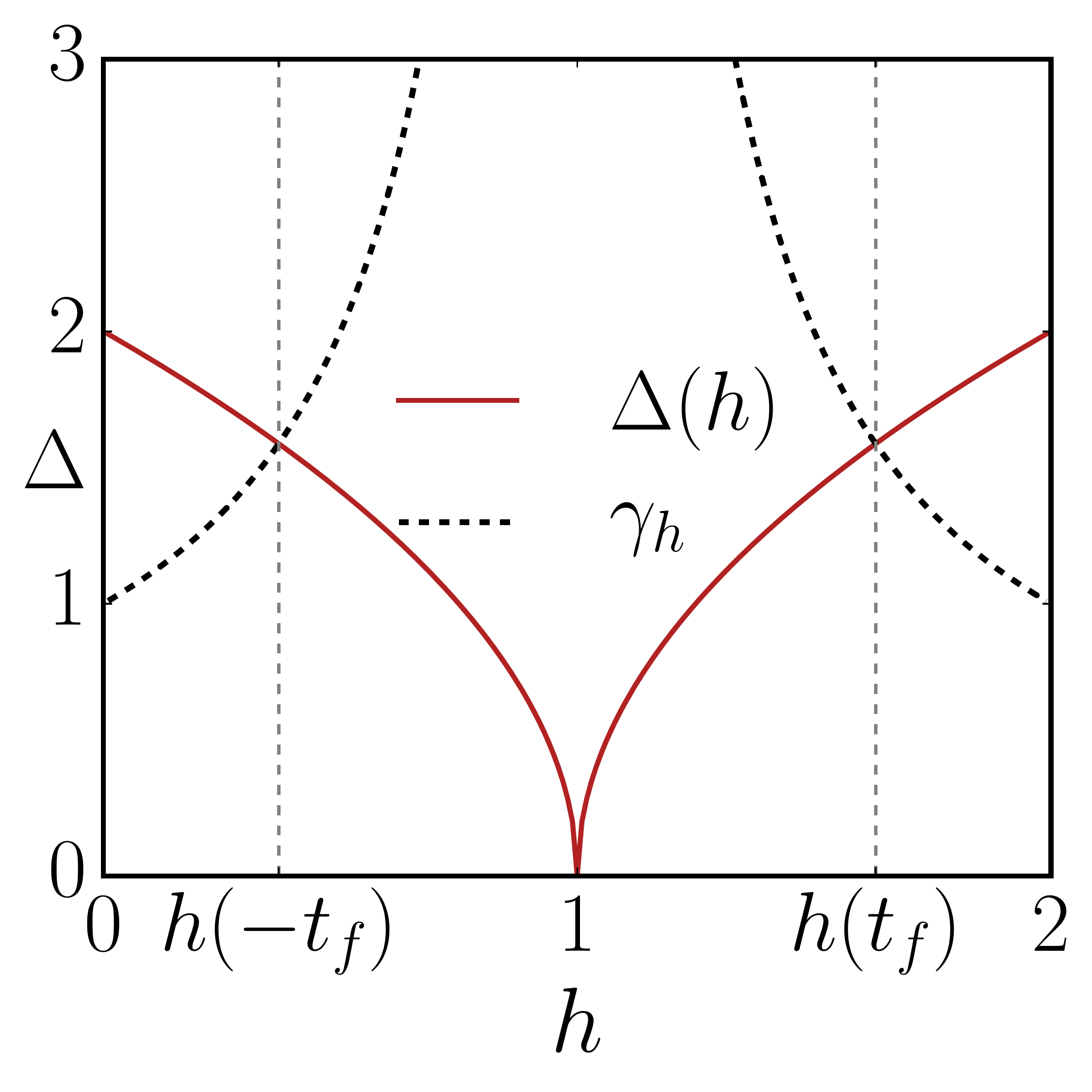}}
\caption{\label{Fig:1}(color online) (a) The dynamics of a spin-$1/2$ chain is analyzed when the strength of the magnetic field $h$ is slowly varied across the paramagnetic-ferromagnetic transition. The lines connecting the sites illustrate that each spin interacts with equal strength $J$ with the rest of the chain. (b) The energy gap $\Delta(h)$ between the ground and the first excited state of the chain is displayed as a function of $h=h_{c}-\delta\cdot t$ (solid red line). The dashed black line shows the rate $\gamma_h=|\dot h|/|h-h_{c}|$ with which the magnetic field is varied in time. The freezing time $t_f$ is defined such that $\gamma_h(\pm t_f)=\Delta(h(\pm t_f))$. 
We determine the scaling with $\delta$ of the quantum heat generated by the quench.}\end{figure}

In this framework, it is believed that the thermodynamics of slow variations (quenches) of control fields across a critical point can be cast in terms of the so-called Kibble-Zurek (KZ) scaling \cite{kibble1976,zurek1985,zurek1996,zurek2005}. The KZ scaling predicts that the heat produced by slow quenches scales with the quench rate as a power law determined by the equilibrium critical exponents \cite{dziarmaga2010, chandran2012, del_campo2014}. This theory has a strong predictive power and has been experimentally tested in a large variety of physical systems 
\cite{schneider2012,ulm2013,pyka2013,mielenz2013,corman2014,braun2015,navon2015,bauerle1996,ruutu1996,carmi1999,monaco2009,weller2008,yukalov2015,chuang1991,ducci1999}. 
However, the validity of the KZ theory seems to be limited to systems where the coherence length diverges with a power law only at the critical point, but is well defined otherwise. 
The KZ hypothesis can be explained as follows: Assume a system of interacting spins in presence of a magnetic field $h$, as illustrated in Fig.~\ref{Fig:1}(a), and let the magnitude of the magnetic field $h$ vary slowly from the paramagnetic to the ferromagnetic phase across the critical point $h_c$. The evolution is adiabatic if the rate of change $\gamma_h=|\dot{h}|/|h-h_{c}|$ is smaller than the energy gap $\Delta(h)$, while in the opposite regime non\-adiabatic effects are expected. Figure \ref{Fig:1}(b) displays the energy gap $\Delta(h)$ as a function of $h$: The gap vanishes as $\Delta(h)\sim |h-h_c|^{z\nu}$ at the critical point, with equilibrium critical exponents $\nu$ and $z$.  The KZ theory identifies the time scale $\pm t_f$ where $\Delta(h)=\gamma_h$ and assumes that in the time window $-t_f<t<t_f$ (namely, when $\Delta(h)<\gamma_h$) the dynamics is frozen. Then the heat $Q$ produced by the quench scales as $Q\sim \xi_f^{-z}$, where $\xi_f\sim |h(t_f)-h_c|^{-\nu}$ is the average size of the spin domains formed at the time $t=-t_f$ in the adiabatic part of the dynamics, yielding $Q\sim |h(t_f)-h_c|^{z\nu}$. For a quench where the magnetic field varies linearly with time as $h=h_c-\delta t$ ($\delta>0$), one obtains \cite{zurek2005,polkovnikov2005,dziarmaga2010,chandran2012}
\begin{equation}
\label{zkz}
Q\sim \delta^{z\nu/(1+z\nu)}.
\end{equation}

Although this separation between adiabatic and frozen (impulse) regime may seem oversimplified, it describes well the behavior found in isolated systems, where the relaxation time is determined by the instantaneous gap between the ground state and the first excited state, and for short-range interactions \cite{silvi2016}. The validity of the KZ scaling \eqref{zkz} has been extensively verified in integrable fermionic systems \cite{dziarmaga2005,damski2005,dutta2017,de_grandi2010,polkovnikov2005}. 
Even at finite temperatures the KZ scaling is a good working hypothesis \cite{biroli2010, tomka2018}. A problem arises, however, when one applies the KZ scaling to systems where the correlation length is ill-defined even away from the critical point \cite{Campa2014,gardas2018}. This is the case for systems with strongly long-ranged interactions where the two-body interaction potential $V(r)\propto r^{-\alpha}$ decays as a power law with the distance $r$, 
with $0\le\alpha<d$ in $d$ spatial dimensions \cite{Campa2014,Kastner2010,Kastner2011}.
It is important to find a paradigm that allows one to extend the KZ scaling hypothesis also to these systems, where very few analytical solutions exist for the out-of-equilibrium dynamics. Moreover, such a paradigm would be important for the development of quantum devices based on quantum annealing, where one aims at preparing many-body quantum states by adiabatic transformations \cite{gardas2018}. Yet, attempts to find the KZ scaling in the Lipkin-Meshkov-Glick (LMG) model \cite{lipkin1965}, a prototype of strong-long-range spin system, have led to seemingly contradicting results~\cite{caneva2008,acevedo2014,hwang2015}. 

In this Letter we derive a scaling theory that encompasses different types of ramps across the critical region of the LMG model. We derive an exact solution which unifies previous findings~\cite{caneva2008,acevedo2014,hwang2015,bachmann2017} and thus provides an important benchmark for numerical studies of the out-of-equilibrium dynamics of quantum strong-long-range systems.

The LMG model describes a system of $N$ spin-$1/2$ degrees of freedom with all-to-all ferromagnetic interactions in a transverse magnetic field $h$, as illustrated in Fig.~\ref{Fig:1}(a). Its Hamiltonian reads 
\begin{align}
\label{h_lmg:0}
H&=-J\Biggl(\frac{1}{N}\sum_{i<j}\sigma^{x}_{i}\sigma^{x}_{j}+h(t)\sum_{i}\sigma^{z}_{i}\Biggr),
\end{align}
where $\sigma_i^{\mu}$ are the Pauli matrices of spin $i$, and the prefactor $1/N$ in front of the interaction term warrants that the energy is extensive \cite{Campa2014}. The parameter $J>0$ scales the energy in units of the interaction strength, and we consider energy and time in units of $J$ and $J^{-1}$, respectively. When the magnetic field $h$ is constant in time, in the thermodynamic limit the LMG model displays equilibrium quantum phase transitions (QPTs) at $h_c=\pm1$ between a symmetric phase ($|h|>h_c$) fully polarized along $x$, and a symmetry-broken phase ($|h|<h_c$) with two degenerate ground states of opposite macroscopic polarization along the $z$ direction. The universality class of the QPT is the same as that of the Dicke model \cite{dicke1954,Hepp1973} and is given by a mean-field theory with critical exponents $z=1/3$ and $z\nu=1/2$ \cite{botet1982, botet1983, das2006}. Experimental realizations include trapped ions \cite{Britton_etal12,Richerme2014,Jurcevic2014} and spinor Bose-Einstein condensates \cite{Ziebold_etal10}. The Dicke model, moreover, has been used to describe quenches in the BEC--BCS crossover regime \cite{Altman2005} as well as the selforganization transition of ultracold bosonic gases in optical cavities \cite{Nagy2010,Baumann2009}.

We now consider a continuous ramp of the control field $h(t)=1-\delta t$, with quench rate $\delta>0$. The protocol starts deep in the paramagnetic phase and can end at the quantum critical point (half-ramp) or 
far in the symmetry-broken phase (full ramp). According to the KZ hypothesis, the quench generates a quantum contribution to the heat with the power-law scaling $Q\sim \delta^{1/3}$, consistent with $z\nu=1/2$ in Eq.~\eqref{zkz}. However, this scaling was not found for full ramps in the numerical studies of Ref.~\cite{caneva2008,acevedo2014}. The existence of a power-law scaling seems inconsistent with the calculation performed in Ref.~\cite{bachmann2017} for a system which is equivalent to the quantum dynamics of the LMG strictly in the thermodynamic limit $N\to\infty$. On the other hand, the scaling $Q\sim \delta^{1/3}$ was reported using an heuristic application of adiabatic perturbation theory for a quench to the critical point \cite{hwang2015}.

\begin{figure*}[t!]
\centering
\subfigure[\large]{\label{Fig2a}\includegraphics[width=.3\textwidth]{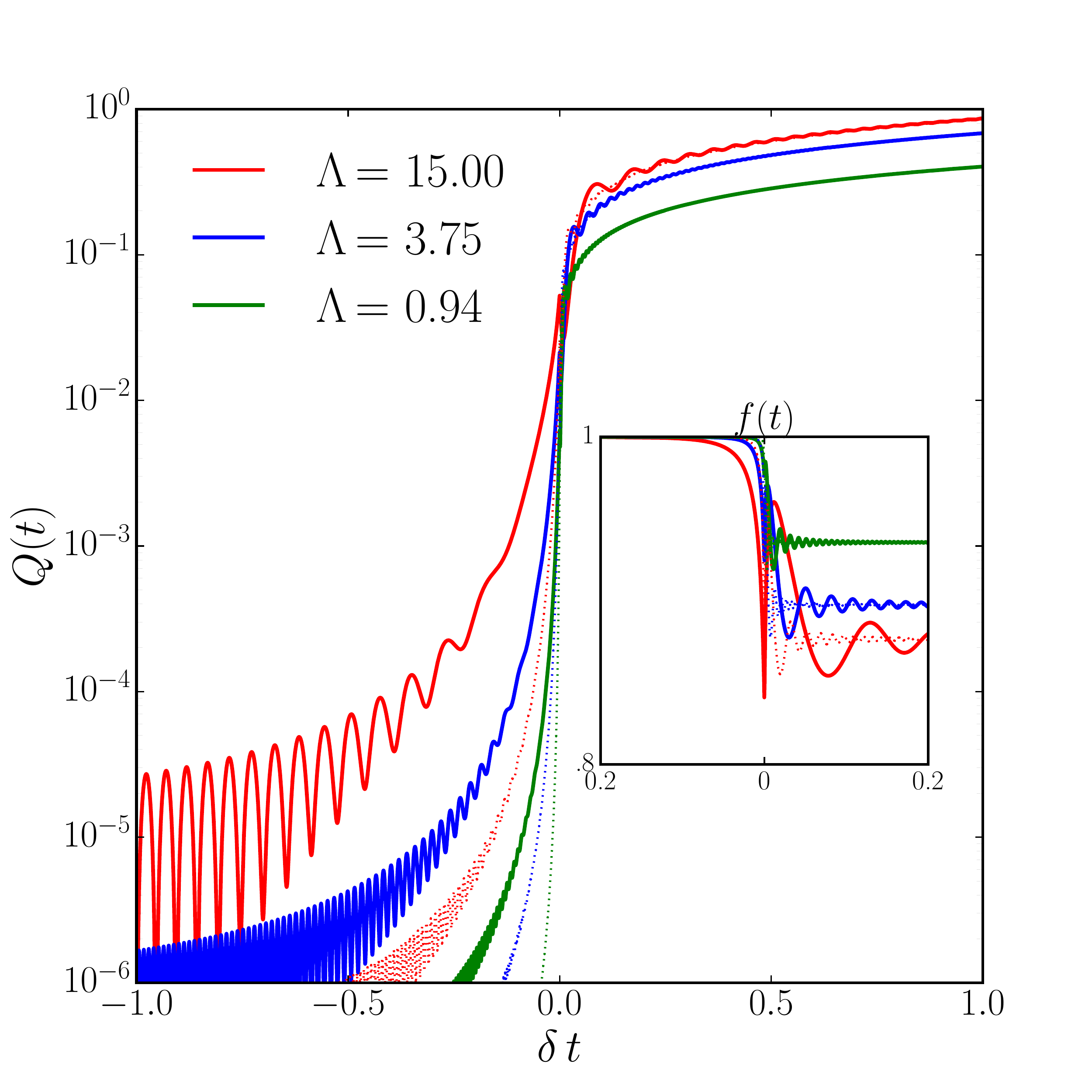}}
\subfigure[\large]{\label{Fig2b}\includegraphics[width=.3\textwidth]{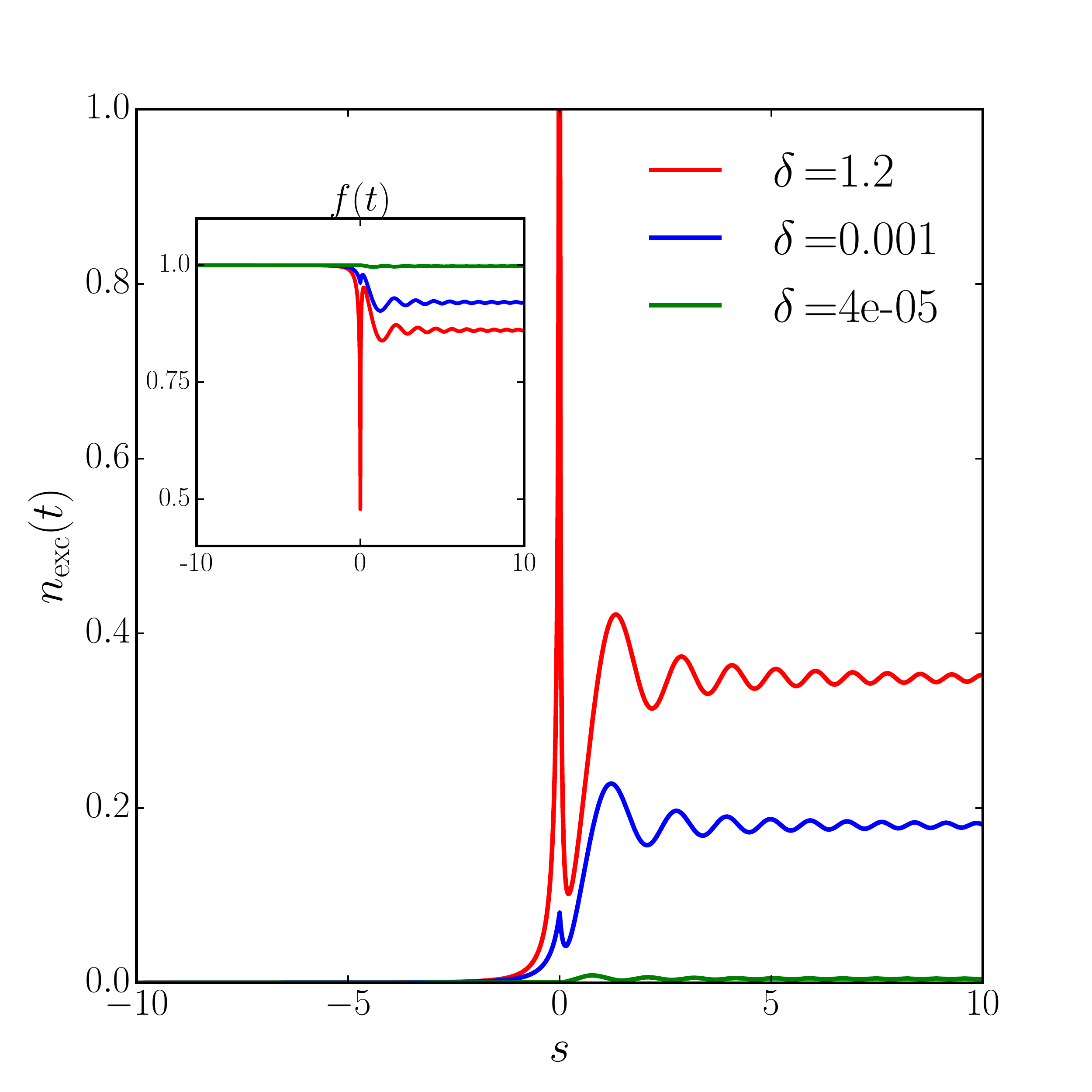}}
\subfigure[\large]{\label{Fig2c}\includegraphics[width=.3\textwidth]{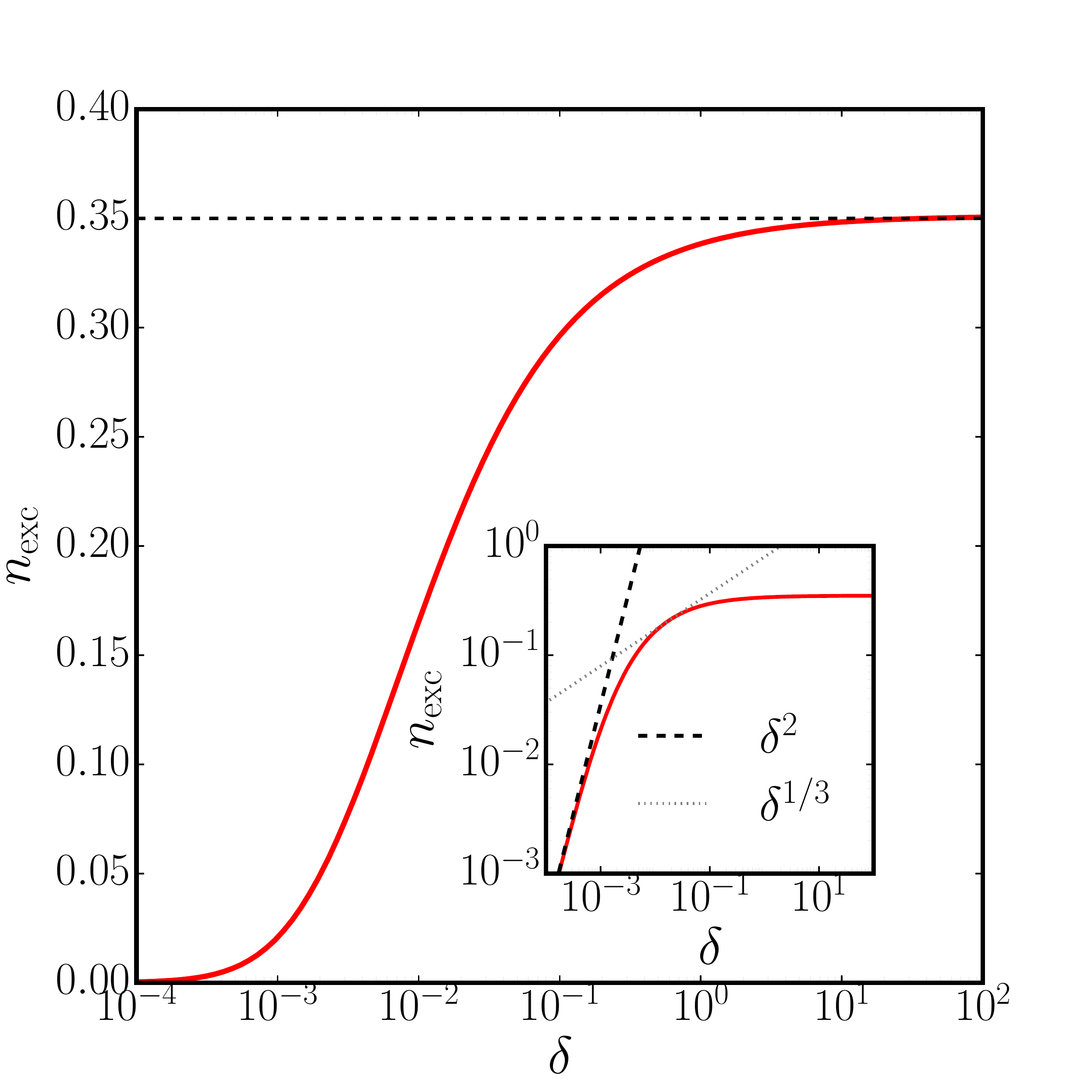}}
\caption{\label{Fig:2} 
(color online) (a) Heat $Q$ generated by the quench, in units of $J$, as a function of time $t$, in units of $1/(J\delta)$. The heat is determined from Eq.~\eqref{n:exc} using Eqs.~\eqref{psi} and \eqref{ermakov_milne}. The plot shows different values of $\Lambda=N\delta$ (indicated in the legend), chosen to be the same as in Ref.~\cite{acevedo2014}. Solid lines correspond to $N=2^9$, dashed lines to $N=2^{12}$. The inset reports the corresponding fidelity $f(t)$. (b) The average number of excitation $n_{\rm exc}(t)$ and the fidelity $f(t)$ are reported as functions of the rescaled time $s=\delta^{1/3}t$ for $N=2^{12}$ and $\delta=4\times10^{-5}$, $10^{-3}$, $1.2$, corresponding to $\Lambda=2\times10^{-1}$, $6$, $6\times10^{3}$, respectively. (c) The number of excitations at the end of the quench, $n_\text{exc}(t_0)$, is reported as a function of $\delta$ for $N=500$ (the behavior for a different system size $N'$ is obtained by rescaling the $\delta$-axis by the factor $N'/N$). The horizontal dashed line indicates the constant value $n_\text{exc}(t_0)=0.35$ of the thermodynamic limit. The inset shows $n_\text{exc}(t_0)$ on a logarithmic scale. The dotted line represents the KZ scaling prediction $\delta^{1/3}$. For the full ramp, there is only an accidental match in the crossover regime and no actual KZ scaling is found.}
\end{figure*}

To explain this puzzle, we solve the Schr\"odinger equation governed by Hamiltonian \eqref{h_lmg:0} for $h(t)=1-\delta t$ with small quench rates $\delta$. For this purpose we rewrite Eq.~\eqref{h_lmg:0} in terms of a single collective spin of length $N$, namely $S_{\mu}=\sum_{i}\sigma^{\mu}_{i}/2$ and $S_{\pm}=S_{x}\pm i\,S_{y}$, such that \cite{dusuel2005}
\begin{align}
\label{h_lmg}
H=-\frac{1}{N}(\boldsymbol{S}^{2}-S_{z}^{2}-N/2)-2h(t)S_{z}-\frac{1}{2N}(S_{+}^{2}+S^{2}_{-}).
\end{align} 
We then perform a $1/N$ expansion around the ground state of the mean-field model \cite{Holstein1940,Auerbach1994}.
The expansion is obtained by first rotating the spin operators to align them with the semiclassical magnetization, which is assumed to adiabatically follow the quench. We then apply a Holstein-Primakoff transformation \cite{Holstein1940} to quadratic order, $S_{z}=N/2-a^{\dagger}a$, $S_{+}=S_{-}^{\dagger}=\sqrt{N}a$, where operators $a$ and $a^\dagger$ satisfy the bosonic commutation relation $[a,a^{\dagger}]=1$. 
Using a Bogoliubov transformation we obtain the diagonal form
\begin{align}
\label{h_hp}
H_{0}=N\,e_{0}(h)+\delta e(h)+\Delta(h)\,b^{\dagger}b
\end{align}
in the new bosonic operators $b$ and $b^{\dagger}$.
Here $e_{0}$ is the thermodynamic mean field energy density, $\delta e$ is a constant mean-field shift, while the quantum fluctuations are described by the quadratic harmonic oscillator term whose frequency is the gap $\Delta$~\cite{dusuel2005}.  
The quantum part of the Hamiltonian \eqref{h_hp} is obtained at leading order in a $1/N$ expansion and is thus strictly valid only in the thermodynamic limit. Nevertheless, by means of the continuous unitary transformation approach~\cite{Wegner2000}, the LMG Hamiltonian can be recast into the form~\eqref{h_hp} also for finite $N$~\cite{dusuel2005,dusuel2004}. Then, the gap reads~\cite{dusuel2005,dusuel2004}
\begin{align}
\Delta=\begin{cases}
2\sqrt{h(h-1)}+\mathcal{F}(N,h) & \text{for $h>1$},\\
2\sqrt{(1-h^{2})}+\mathcal{F}(N,h) & \text{for $h<1$},
\end{cases}
\end{align}
where for large $N$ and $h\neq 1$ the function $\mathcal{F}(N,h)\sim1/N$, while at the critical point the gap scales as $\Delta \propto 1/N^{1/3}$ \cite{dusuel2005,dusuel2004}.
The Hamiltonian \eqref{h_hp} corresponds to a single harmonic oscillator with time-dependent frequency $\Omega(t)=\Delta(h(t))$, which can be exactly solved in terms of the dynamical basis
\begin{align}
\label{psi}
\psi_{n}(x,t)=&\left(\frac{e^{-i4\phi(t)}}{2\pi\xi^{2}(t)}\right)^{\frac{1}{4}}\frac{e^{-\tilde{\Omega}(t)x^2/2}}{\sqrt{2^{n}n!}}
H_{n}\left(\frac{x}{\sqrt{2}\xi(t)}\right).
\end{align}
Here, $H_{n}$ is the Hermite polynomial of degree $n$, $\phi(t)$ is a phase factor, $\tilde{\Omega}(t)=1/2\xi^{2}+i\dot{\xi}/\xi$ is the effective frequency and $\xi(t)$ is a time dependent scale factor which obeys the Ermakov-Milne equation \cite{leach2008,milne1930,pinney1950}
\begin{align}
\label{ermakov_milne}
\ddot{\xi}(t)+\Omega(t)^{2}\xi(t)=\frac{1}{4\xi(t)^{3}}.
\end{align}
The wave function evolves from the time $t=-t_0=-1/\delta$ until the final time $t=t_0$ (full ramp) or $t=0$ (half ramp). In the adiabatic limit the $\psi_n(x,t)$ coincide with the instantaneous eigenstates $\psi_{n}^{{\rm ad}}(x,t)$ of Hamiltonian $H(t)$, which are the solutions of Eq.~\eqref{psi} after setting $\dot{\xi}=\ddot{\xi}=0$ in Eq.~\eqref{ermakov_milne} and thus $\xi(t)^2=1/(\sqrt{2}\Omega(t))$ in Eq.~\eqref{psi}. We denote the overlap integral between $\psi_0(x,t)$ and the eigenfunctions $\psi_{n}^{{\rm ad}}(x,t)$ of the adiabatic basis by $c_n(t)=\int \psi_{n}^{{\rm ad}*}(x,t)\psi_{0}(x,t)dx$. Its explicit expression is derived in App.\,\ref{AppA} and in Ref.~\cite{Dabrowski2016}. By means of the fidelity $f(t)=|c_{0}(t)|^2$ we verify that the initial state $\psi_0(x,-t_0)$ coincides with the ground state of the Hamiltonian $H$ at $h=0$ (see Fig.~\ref{Fig:2}(a) inset). The heat generated at time $t>-t_0$ is proportional to the excitation number $n_{\rm exc}(t)$, $Q(t)=\Omega(t)n_\text{exc}(t)$, where
\begin{equation}
\label{n:exc}
n_\text{exc}(t) = \sum_{n=1}^\infty n |c_n(t)|^2.
\end{equation}

Figure~\ref{Fig:2}(a) displays the time evolution of the heat $Q$ obtained from Eqs.~\eqref{psi} and \eqref{ermakov_milne} for the parameters of Ref.~\cite{acevedo2014}. The curves in Fig.~\ref{Fig:2}(a) reproduce the ones found numerically in Ref.~\cite{acevedo2014} by direct numerical computation of the dynamics of $2^9-2^{11}$ spins with Hamiltonian \eqref{h_lmg:0}. As in Ref.~\cite{acevedo2014}, we observe a drop of the fidelity (inset) at the critical point, indicating the loss of adiabaticity. For the chosen parameter values, however, the dynamics remains close to adiabatic with fidelity $f>80\%$. No numerical evidence of KZ scaling was found in Ref.~\cite{acevedo2014}, and it was conjectured that this may be due to universal finite-size scaling functions at the critical point. We now analyze the scaling of the gap and show that it depends only on the dimensionless parameter
\begin{equation}
\label{omega:0}
\Lambda=N\delta\,.
\end{equation}
To this end, we approximate the oscillator frequency as $\Omega(t)^{2}=-4\delta t+1/N^{2z_{\mathrm{eff}}}$ for  $t<0$ and $\Omega(t)^{2}=8\delta t+1/N^{2z_{\mathrm{eff}}}$ for $t\geq0$, up to nonuniversal intensive factors. We have verified numerically that further terms are irrelevant as they become subleading in the critical stage ($t\simeq 0$) of the dynamics. The overall effect of finite-size fluctuations is captured by an effective finite-size scaling exponent $z_\text{eff}$ in the range $1/3<z_\text{eff}<1$. The full numerical solution of Eq.~\eqref{ermakov_milne} indicates that finite-size corrections only become important at $t\simeq0$, hence we assume $z_\text{eff}=1/3$ for the purpose of our discussion. We identify the scaling relations by performing the transformation $\xi=\delta^{-1/6}\tilde{\xi}$ and $t=\delta^{-1/3}s$ (note that, apart from a factor $\Lambda^{2/3}$, $s$ is the same rescaled time variable as in Fig.~2(c) of Ref.~\cite{acevedo2014}). This transformation leads to the Schr\"odinger equation of a quantum harmonic oscillator with effective frequency $\Omega(s)$, where
\begin{align}
\Omega(s)^{2}=\begin{cases}
-4s+\Lambda^{-2/3} & \text{for $s<0$},\\
8s+\Lambda^{-2/3} & \text{for $s>0$}.
\end{cases}
\end{align}
Hence, $\Lambda$ is now the sole physical parameter which encodes the quench rate $\delta$ and the only scale which determines the dynamical behavior at the critical point. We can identify two asymptotic regimes: (i) the limit $\Lambda\ll 1$, where the quench rate is much smaller than the gap and thus the dynamics is expected to be adiabatic. This regime is expected to provide the Landau-Zener scaling $n_{\rm exc}\sim \delta^2$, and corrections to adiabaticity scale with $\Lambda^{2}$~\cite{damski2005, dziarmaga2010, polkovnikov2005, de_grandi2010}. (ii) For $\Lambda\gg 1$, instead, the system approaches the thermodynamic limit where the dynamics is independent of $\Lambda$ to leading order in $1/\Lambda$. In this limit, therefore, excitations and fidelity are expected to be independent of $\delta$. This result is consistent with the prediction of Ref.~\cite{bachmann2017} for a slow quench of the frequency of a single harmonic oscillator, albeit with a different power law in time. 

Figure \ref{Fig:2}(b) shows the time evolution of $f$ and $n_\text{exc}$ for values of $\Lambda$ in the two asymptotic regimes as well as in the intermediate regime. The final value $n_{\rm exc}(t_0)$, which we extract from these calculations, is reported in Fig.~\ref{Fig:2}(c) as a function of $\delta$ for fixed $N$. We observe the Landau-Zener scaling $n_\text{exc}\sim \delta^2$ for $\delta\ll 1/N$, in agreement with our scaling arguments. For $\delta\gg 1/N$ the excitation number tends to a constant value, which we obtain in the thermodynamic limit as $n_{\text{exc},\infty}\approx0.35$. Even though there is no power-law scaling in the thermodynamic limit, the final number of defects $n_\text{exc}(t_0)$ still depends on the scaling of the gap at $s\to 0$. This number is therefore universal and hints at a connection between out-of-equilibrium dynamics and universal equilibrium properties. Since the slope of the curve $n_\text{exc}$ varies continuously as a function of $\delta$, it contains also an interval of values with scaling $\delta^{1/3}$. This scaling, which would agree with the KZ prediction, is clearly only a crossover. 

A very different result is instead found for a half ramp which starts or ends at the quantum critical point (QCP). As we show in the App.\,\ref{SM3}, in that case $n_{\rm exc}\sim \Lambda^{1/3}$. If the ramp ends exactly at the QCP where the gap scales as $1/N^{1/3}$, the heat scales as $Q\sim \delta^{1/3}$ also in the thermodynamic limit independent of system size. This result is in agreement with the predictions of Refs.~\cite{polkovnikov2005,hwang2015} and the KZ hypothesis. However, it occurs only for a half ramp ending exactly at the QCP and thus depends sensitively on the endpoint. For any other quench, on the contrary, $Q$ exhibits a functional dependence on $\Lambda$ (and thus, for $N$ fixed, of $\delta$) which can be reduced to a power law only for finite systems in the adiabatic, Landau-Zener limit. This behavior is markedly different from the one found in short-range interacting systems. It shows that the hypothesis of an impulse regime, where the system is expected to freeze in the time interval when the gap is smaller than the quench rate, $t\in[-t_f,t_f]$ of Fig.~\ref{Fig:1}, does not hold for the LMG model, and thus strictly speaking the KZ scaling does not apply. These results are also valid for the Dicke model, whose finite-size
corrections to the gap have the same scaling with $N$ \cite{vidal2006}. Moreover, the LMG model is the limit $\alpha\to 0$ of a spin chain with general power-law interactions $1/r^\alpha$, while the zero order $1/N$ expansion corresponds to the leading order spin-wave approximation \cite{Ruckriegel2012,Lerose2018}. From preliminary calculations we expect our conclusions to hold for any power law with $0\le \alpha<1$ in one dimension, i.e., when the correlation length is ill-defined. 

Our analytical theory describes quantum contributions to the heat.  These are due to excitations on top of the mean-field spin, and are therefore valid when the non-adiabatic corrections of the mean-field energy are smaller than the quantum heat. Assuming that the semiclassical evolution is analytical in $\delta$ and that, in a cyclic process, no work is done on the system in the adiabatic limit $\delta\to 0$, the semiclassical contribution to the heat would scale as $N\delta^2$ \cite{zwerger2008}. In this perspective the classical motion follows adiabatically the drive and the quench dynamics is independent of the quench direction\,\cite{footnote}. The nonadiabatic quantum contribution dominates the dynamics for $\delta\lesssim N^{-2/5}$, at least for the half ramp.
These observations also suggest that the scaling $N\delta^2$ found for slower quenches \cite{caneva2008} is dominated by mean-field dynamics, where the quantum contribution to the heat is not yet visible. Our predictions can be experimentally tested in assemblies of trapped ions with all-to-all interactions \cite{Britton_etal12,zhang2017,Safavi2018} and in spinor Bose-Einstein condensates \cite{Ziebold_etal10}. The regime corresponding to $\Lambda\gg1$, where the quantum residual energy tends to a constant, could be observed in simple systems such as the Rabi model \cite{hwang2015}. Finally, our study applies to quenches of lasers illuminating ultracold atoms in cavity quantum electrodynamics \cite{Baumann2009,klinder2015, keller2018,Hruby2018} where photons mediate all-to-all interactions \cite{Ritsch2013,Vaidya2018}. 

\acknowledgments N.\,D.\ acknowledges fruitful discussions with G.~Gori, G.\,M.~Graaf, S.~Ruffo, and A.~Trombettoni.  G.\,M.\ is grateful to L.~Hruby, S.~J\"ager, H.~Ritsch, and V.~Torggler for stimulating discussions. Financial support by the DFG Collaborative Research Centre ``SFB 1225 ISOQUANT'', by the DFG DACH project ``Quantum crystals of matter and light'', by the German Ministry of Education and Research (BMBF) via the Quantera project ``NAQUAS'', and by the Competitive Programme for Rated Researchers of the NRF South Africa is acknowledged.

\textit{Note added.} After the submission of this paper, a preprint by Ming Xue, Shuai Yin and Li You
\cite{Xue2018} appeared on arXiv.  The authors describe the universal dynamics across the quantum critical 
point of a ferromagnetic spinor atomic Bose-Einstein condensates, whose universal behavior is equivalent to the one of
the LMG model. Our analytical predictions agree with these numerical results for finite system size.
\onecolumngrid
\appendix
\section{Exact solution and Defect density}
\label{AppA}
The dynamics of the quantum harmonic oscillator can be solved exactly \cite{Lewis1967,Lewis1968,Lewis1969} and any dynamical state $\psi(x,t)$ in the representation of the coordinate $x$ can be expressed as
\begin{align}
\label{dyn_exp}
\psi(x,t)=\sum \alpha_{n}\psi_{n}(x,t),
\end{align}
where $\alpha_{n}$ are time independent constants and the dynamical eigenstates are given by
\begin{align}
\label{Dyn_Eigen}
\psi_{n}(x,t)=\frac{1}{\sqrt{2^{n}n!}}\left(\frac{1}{2\pi\xi^{2}(t)}\right)^{\frac{1}{4}}e^{-\tilde{\Omega}(t)\frac{x^{2}}{2}}
H_{n}\left(\frac{x}{\sqrt{2}\xi(t)}\right)e^{-i\left(n+\frac{1}{2}\right)\lambda(t)}.
\end{align}
The effective frequency $\tilde{\Omega}(t)$ can be expressed in terms of the effective width $\xi(t)$ as
\begin{align}
\tilde{\Omega}(t)=-i\frac{\dot{\xi}(t)}{\xi(t)}+\frac{1}{2\xi^{2}(t)},
\end{align}
and
\begin{align}
\lambda(t)=\int^{t}\frac{dt'}{2\xi^{2}(t')}
\end{align}
is the total phase accumulated at time $t$. The exact time evolution of each harmonic oscillator is then fully described by a single complex parameter, which is the effective width $\xi(t)$ satisfying the Ermakov-Milne equation
\begin{align}
\label{ermakov_eq}
\ddot{\xi}(t)+\Omega(t)^{2}\xi(t)=\frac{1}{4\xi^{3}(t)}.
\end{align}
If the initial state is a pure state of the basis \eqref{Dyn_Eigen}, say, the ground state, then all the coefficients $\alpha_{n}$ of Eq. \eqref{dyn_exp} vanish except for the coefficient $\alpha_0$. This holds also at all later times, and thus in the exact dynamical basis \eqref{Dyn_Eigen} no excited states will be generated. However at each time $t>t_{0}$ the dynamical pure state $\psi_{0}(x,t)$ will, in general, be different from the instantaneous equilibrium ground state since the effective width will not coincide with the equilibrium basis $\psi_{n}^{\rm ad}(x,t)$, whose wave functions read 
\begin{align}
\label{Eq_Eigen}
\psi_{n}^{\rm ad}(x,t)=\frac{1}{\sqrt{2^{n}n!}}\left(\frac{\Omega(t)}{\pi}\right)^{\frac{1}{4}}e^{-\Omega(t)\frac{x^{2}}{2}}H_{n}\left(x\sqrt{\Omega(t)}\right).
\end{align}

Therefore, if we decompose any pure state $\psi_{n}(x,t)$ of the dynamical basis using the instantaneous equilibrium basis, it will generally contain a number of equilibrium excitations. Then, assuming the evolution to start in the ground state at $t=-t_0$, the number of excitations in the instantaneous equilibrium basis at time $t$ is given by~\cite{Dabrowski2016}
\begin{align}
\label{Exc_N}
n_\text{exc}(t)=\sum_{n\in 2\mathbb{N}}n|c_{n}(t)|^{2},
\end{align} 
where the coefficients
\begin{align}
\label{trans_amp}
c_{n}(t)=\int_{-\infty}^{+\infty}dx\,\psi_{n}^{*}(x,t)\psi_{0}(x,t)
\end{align}
are the transition amplitudes between the dynamical state and the instantaneous equilibrium basis.
The definition \eqref{Exc_N} can be evaluated by choosing different basis sets for the evaluation of transition amplitudes rather than the eigenstates given in \eqref{Eq_Eigen} \cite{Dabrowski2016}. However, the basis of the eigenstates  in \eqref{Eq_Eigen} is the most natural choice in the context of the Kibble-Zurek mechanism.

Using definition \eqref{Exc_N} together with Eq.~\eqref{trans_amp} one can derive an explicit expression for the number of excitations $n_\text{exc}(t)$. To this aim we evaluate the transition amplitudes
\begin{align}
c_{n}(t)=&\int_{-\infty}^{+\infty}dx\psi_{n}^{\rm ad *}(x,t)\psi_{0}(x,t)
=\frac{1}{\sqrt{2^{n}n!\pi}}\left(\frac{\Omega(t)}{2\xi^{2}(t)}\right)^{\frac{1}{4}}\int_{-\infty}^{+\infty}dx e^{-(\Omega(t)+\tilde{\Omega}(t))\frac{x^{2}}{2}}H_{n}\left(\sqrt{\Omega(t)}x\right).
\end{align}
Performing a change of variable the integral can be cast into the form
\begin{align*}
\int_{-\infty}^{+\infty}dx e^{-(\Omega(t)+\tilde{\Omega}(t))x^{2}}H_{n}\left(\sqrt{\tilde{\omega(t)}}x\right)=(\Omega(t))^{-\frac{1}{2}}\int_{-\infty}^{+\infty} e^{-\left(\frac{\tilde{\Omega}(t)}{\Omega(t)}+1\right)\frac{s^{2}}{2}}H_{n}\left(s\right)ds.
\end{align*}
Next we employ the generating function for Hermite polynomials in the integral,
\begin{multline}
\int_{-\infty}^{+\infty} e^{-\left(\frac{\tilde{\Omega}(t)}{\Omega(t)}+1\right)\frac{s^{2}}{2}}H_{n}\left(s\right)ds=\lim_{t\to0}\frac{d^{n}}{dt^{n}}\int_{-\infty}^{+\infty} e^{-\left(\frac{\tilde{\Omega}(t)}{\Omega(t)}+1\right)\frac{s^{2}}{2}}e^{2st-t^{2}}ds=\sqrt{\frac{2\pi}{\left(\frac{\tilde{\Omega}(t)}{\Omega(t)}+1\right)}}\lim_{t\to0}\frac{d^{n}}{dt^{n}}e^{-t^{2}\frac{\left(\tilde{\Omega}(t)-\Omega(t)\right)}{\left(\Omega(t)+\tilde{\Omega}(t)\right)}}\\
=\begin{cases}
\displaystyle\sqrt{\frac{2\pi}{\left(\frac{\tilde{\Omega}(t)}{\Omega(t)}+1\right)}}\frac{n!}{\frac{n}{2}!}\left(\frac{\tilde{\Omega}(t)-\Omega(t)}{\tilde{\Omega}(t)+\Omega(t)}\right)^{n/2} & \text{for $n\in 2\mathbb{Z}$},\\
0 & \text{for $n\in 2\mathbb{Z}+1$}.
\end{cases}
\end{multline}
Thus the probability of having $n$ excitations in the evolved state at the time $t$ is given by
\begin{align}
|c_{n0}(t)|^{2}=\frac{(n-1)!!}{n!!}\frac{\sqrt{2\Omega(t)}}{\xi(t)\left|\tilde{\Omega}(t)+\Omega(t)\right|}\left|\frac{\tilde{\Omega}(t)-\Omega(t)}{\tilde{\Omega}(t)+\Omega(t)}\right|^{n}.
\end{align}
We insert this expression into Eq.~\eqref{Exc_N} and obtain the number of excitations at time $t$,
\begin{align}
\label{expl_exc_num}
n_{exc}(t)=\frac{\xi^{2}}{2\Omega(t)}\left[\left(\frac{1}{2\xi^{2}}-\Omega(t)\right)^{2}+\left(\frac{\dot{\xi}}{\xi}\right)^{2}\right].
\end{align}
From this expression we can also determine the ground state fidelity $f(t)=|c_0(t)|^2$, which reads
\begin{align}
f(t)=|c_{00}(t)|^{2}=\frac{2\Omega(t)}{\xi(t)}\left[\left(\frac{1}{2\xi^{2}}-\Omega(t)\right)^{2}+\left(\frac{\dot{\xi}}{\xi}\right)^{2}\right]^{-1/2}.
\end{align}

\section{The slow quench from the critical point}
\label{SM2}

Here we consider the half ramp $h=1-\delta t$ with $t\in[0,t_0]$, where the magnetic field is quenched from the critical point far into the symmetry broken phase. For $t>0$ the harmonic oscillator frequency varies as  
\begin{align}
\Omega(t)^{2}=\delta t+1/N^{2/3}
\end{align}
where, once again, we discarded time-dependent finite-size corrections and sub-leading terms which do not modify the universal behavior as well as unimportant numerical factors.
It is convenient to employ the rescaling
\begin{align}
\label{resc}
\xi=\delta^{-1/6}\tilde{\xi},\qquad t=\delta^{-1/3}s
\end{align}
already introduced in the main text. The Ermakov-Milne equation now reads
\begin{align}
\label{ermakov_eq_app}
\ddot{\xi}(s)+\Omega(s)^{2}\xi(t)=\frac{1}{4\xi^{3}(s)},
\end{align}
where the rescaled frequency is given by
\begin{align}
\label{eq_rescaling}
\tilde{\Omega}(s)^{2}=s+\Lambda^{-2/3}
\end{align}
with $\Lambda=N\delta$. From now one we will suppress the \~ superscript over rescaled quantities, as they are the only ones appearing in the following. The solution of Eq.~\eqref{ermakov_eq_app} can be constructed from that of the associated classical harmonic oscillator
\begin{align}
\label{cl_h_osc}
\ddot{x}(s)+\Omega(s)^{2}x(s)=0.
\end{align}
This equation admits the two independent solutions
\begin{align}
x_{1}(s)=\mathrm{Ai}\left(-\Omega^{2}(s)\right),\qquad
x_{2}(s)=\mathrm{Bi}\left(-\Omega^{2}(s)\right)
\end{align}
in terms of the Airy functions $\mathrm{Ai}$ and $\mathrm{Bi}$. The two functions $x_{1}(s)$ and $x_{2}(s)$ have the constant and finite Wronskian
\begin{align}
\mathrm{Wr}(x_{1},x_{2})=\frac{1}{\pi}.
\end{align}
It is convenient to rewrite the solutions of Eq.~\eqref{ermakov_eq_app} as a pair of complex conjugate solutions $w$ and $w^*$ with
\begin{align}
\label{sol_def}
w=a x_{1}(s)+b x_{2}(s),
\end{align}
where $a\in\mathbb{C}$ and $b\in\mathbb{R}$ are constants. Since Eq.~\eqref{cl_h_osc} is homogeneous one can rescale the two solution by a constant factor and subsequently, without loss of generality, impose $b=1$. The function
\begin{align}
\label{xi_sol}
\xi(s)=\sqrt{ww^*}
\end{align}
is a solution of the Ermakov-Milne equation \eqref{ermakov_eq} if
\begin{align}
\label{wronsk_condition}
\mathrm{Wr}(w,w^{*})=2i\mathrm{Im}(a)\mathrm{Wr}(x_{1},x_{2})=i,
\end{align}
which uniquely fixes the imaginary part of $a$. To completely define the solution, it is required to find the appropriate value of $\mathrm{Re}(b)$ which satisfies the boundary condition
\begin{align}
\frac{1}{2\xi(0)^{2}}=\Lambda^{-1/3}.
\end{align}
By inverting this expression one readily obtains 
\begin{align}
\label{cond2}
2\xi(0)^{2}=\Lambda^{1/3}.
\end{align}
When the thermodynamic limit is taken first, then the right-hand side of this expression diverges. We consider the limit where $\Lambda^{-2/3}$ can be neglected in the argument of the Airy functions, but the right-hand side of Eq.~\eqref{cond2} remains finite, leading to the expression
\begin{align}
\mathrm{Re}(a)=\frac{x_{2}(0)}{x_{1}(0)}\pm\sqrt{\frac{\Lambda^{1/3}}{2x_{1}(0)}+\mathrm{Im}(a)}.
\end{align}
Hence, $\mathrm{Re}(a)$ diverges in the thermodynamic limit for finite values of the ramp rate $\delta$. Making use of the asymptotic expression for the Airy functions and neglecting oscillatory terms as it was done in Ref.~\cite{polkovnikov2008}, the asymptotic time limit of the scale parameter $\xi$ is
\begin{align}
\lim_{s\to\infty}\xi(s)^{2}=\frac{1+|a|^{2}+2\mathrm{Re}(a)}{\pi\Omega(s)}
\end{align}
which, inserted into Eq.~\,\eqref{expl_exc_num}, leads to
\begin{align}
n_{\mathrm{exc}}(t)\propto \mathrm{Re}(a)^{2},
\end{align}
which scales as $\Lambda^{1/3}$. This expression is equivalent to the one obtained in~\cite{polkovnikov2008} for a gapless system and suggests a KZ scaling for large sizes. 

\section{The slow quench to the critical point}
\label{SM3}
Here we consider the case of a half ramp $h=h_{c}-\delta\,t$ with $t\in[-t_0,0]$, starting in the paramagnetic phase and ending at the critical point. The solution to Eq.~\eqref{ermakov_eq_app} is still given by Eqs.~\eqref{sol_def} and \eqref{xi_sol}, but with the boundary conditions
\begin{align}
\label{b_cond}
\lim_{s\to-\infty}\frac{1}{2\xi(s)^{2}}=\Omega(s),\qquad
\lim_{s\to-\infty}\dot{\xi}(s)=0.
\end{align}
These conditions are consistent with the system being in the adiabatic ground state at large $|t|$. In the $s\to\infty$ limit, $\Omega^{2}$ diverges and one must use the asymptotic expansion for the Airy functions
\begin{align}
\lim_{s\to-\infty}x_{1}(s)\sim\frac{\cos\left(\frac{2}{3}\Omega^{3}-\frac{\pi}{4}\right)}{\sqrt{\pi}\Omega^{1/4}},\qquad
\lim_{s\to-\infty}x_{2}(s)\sim \frac{\sin\left(\frac{2}{3}\Omega^{3}-\frac{\pi}{4}\right)}{\sqrt{\pi}\Omega^{1/4}}.
\end{align}
In order to satsify \eqref{b_cond}, the oscillatory terms in the expression for $\xi$ must cancel for large $s$, implying
\begin{align}
\mathrm{Re}(a)=0,\qquad
\mathrm{Im}(a)=b.
\end{align}
Moreover one has to impose the condition 
\begin{align}
\label{wronsk_condition}
\mathrm{Wr}(w,w^{*})=2i\mathrm{Im}(a)b\mathrm{Wr}(x_{1},x_{2})=i,
\end{align}
which fully determines the coefficients in Eq.~\eqref{sol_def},
\begin{align}
\mathrm{Im}(a)=b=\sqrt{\frac{\pi}{2}}.
\end{align}
The resulting expression for the scale factor is
\begin{align}
\xi(s)^{2}=\frac{\pi}{2}\mathrm{Ai}\left(-\Omega(s)^{2}\right)^{2}+\frac{\pi}{2}\mathrm{Bi}\left(-\Omega(s)^{2}\right)^{2},
\end{align}
and the number of defects is given by the formula
\begin{align}
\label{expl_exc_num_resc}
n_{exc}(s)=\frac{\xi(s)^{2}}{2\Omega(s)}\left[\left(\frac{1}{2\xi(s)^{2}}-\Omega(s)\right)^{2}+\left(\frac{\dot{\xi(s)}}{\xi(s)}\right)^{2}\right],
\end{align}
which is identical to Eq.~\eqref{expl_exc_num}, since this quantity is invariant under the rescaling in Eq.~\eqref{resc}. The number of defects at the final point of the ramp (which is the critical point) is obtained by evaluating Eq.~\eqref{expl_exc_num_resc} at $s=0$. At this instant the rescaled frequency is given by its finite-size correction $\Omega(0)=\Lambda^{-1/3}$, and the scale factor reads
\begin{align}
\xi(0)^{2}=\frac{\pi}{2}\mathrm{Ai}\left(-\Lambda^{-2/3}\right)^{2}+\frac{\pi}{2}\mathrm{Bi}\left(-\Lambda^{-2/3}\right)^{2}.
\end{align}
Let us consider the thermodynamic limit $\Lambda\to \infty$ first. In this case the argument of the Airy functions goes to zero and the terms in the square brackets of Eq.~\eqref{expl_exc_num_resc} read
\begin{align}
\frac{1}{4\xi(0)^{4}}=\frac{3^{8/3}\Gamma(2/3)^{4}}{16\pi^{2}},\qquad
\left(\frac{\dot{\xi}(0)}{\xi(0)}\right)=\frac{3^{2/3}\Gamma(2/3)^{2}}{\Gamma(1/3)^{2}},
\end{align}
which, inserted into the defect density \eqref{expl_exc_num_resc}, leads to
\begin{align}
n_{\mathrm{exc}}(0)=\frac{\pi\,\Lambda^{1/3}}{3^{2/3}\Gamma(1/3)^{2}},
\end{align}
where we restricted to the leading term in the $\Lambda\to\infty$ limit. Therefore the result for the number of excitations diverges in the thermodynamic limit with a power $N^{1/3}$. However the residual heat is finite since it is obtained by multiplying the divergent defect density with the vanishing oscillator frequency $Q(0)=\Delta(0)n_{\mathrm{exc}}(0)$, leading to
\begin{align}
Q=\frac{\pi\,\delta^{1/3}}{3^{2/3}\Gamma(1/3)^{2}},
\end{align}
which agrees with the KZ scaling of Ref.~\cite{hwang2015}. For a finite-size system $N<\infty$, the slow ramp limit $\delta\to 0$ coincides with the $\Lambda\to 0$ limit of Eq.~\eqref{expl_exc_num_resc} evaluated at $s=0$. The leading term in this case is generated by the velocity correction to the effective frequency
\begin{align}
\lim_{\Lambda\to 0}\frac{\dot{\xi}(0)}{\xi(0)}=-\frac{5}{24}\Lambda^{2/3},
\end{align}
which, substituted into Eq.~\eqref{expl_exc_num_resc} evaluated at $s=0$, gives
\begin{align}
n_{\mathrm{exc}}(0)=\frac{25}{2304}\Lambda^{2}\propto \delta^{2},
\end{align}
which leads to the expected adiabatic correction for the residual heat $Q\propto\delta^2$ in a finite-size system~\cite{polkovnikov2008,zwerger2008}.
\twocolumngrid

\bibliographystyle{apsrev_cm}
\bibliography{LR_KZM}

\begin{thebibliography}{78}
\bibitem{footnote}
For direct quenches from the paramagnetic to the ferromagnetic phase an
  infinitesimal symmetry breaking field has to be considered in explicit
  calculations.
\end{thebibliography}
\end{document}